\documentclass[twocolumn,prb,letterpaper,superscriptaddress]{revtex4}

\usepackage{amsmath}
\usepackage{amssymb}
\usepackage{braket}
\usepackage{graphicx}
\usepackage{verbatim}
\usepackage{color}
\usepackage{amsthm}
\usepackage{bm}
\usepackage{epsfig}
\usepackage{subfigure}

\newcommand{\be}{\begin{equation}}
\newcommand{\ee}{\end{equation}}
\def \be{\begin{equation}}
\def \ee{\end{equation}}
\def \ba{\begin{array}}
\def \ea{\end{array}}
\def \bea{\begin{eqnarray}}
\def \eea{\end{eqnarray}}

\begin{document}

\title{Magneto-Josephson effects and Majorana bound states in quantum wires}

\author{Falko Pientka}
\affiliation{\mbox{Dahlem Center for Complex Quantum Systems and Fachbereich Physik, Freie
Universit\"at Berlin, 14195 Berlin, Germany}}

\author{Liang Jiang}
\affiliation{Department of Applied Physics, Yale University, New Haven, Connecticut 06511, USA}

\author{David Pekker}
\affiliation{Department of Physics, California Institute of Technology, Pasadena,
California 91125, USA}

\author{Jason Alicea}
\affiliation{Department of Physics, California Institute of Technology, Pasadena,
California 91125, USA}

\author{Gil Refael}
\affiliation{\mbox{Dahlem Center for Complex Quantum Systems and Fachbereich Physik, Freie
Universit\"at Berlin, 14195 Berlin, Germany}}
\affiliation{Department of Physics, California Institute of Technology, Pasadena,
California 91125, USA}

\author{Yuval Oreg}
\affiliation{Department of Condensed Matter Physics, Weizmann Institute of Science,
Rehovot, 76100, Israel}

\author{Felix von Oppen}
\affiliation{\mbox{Dahlem Center for Complex Quantum Systems and Fachbereich Physik, Freie
Universit\"at Berlin, 14195 Berlin, Germany}}
\affiliation{Department of Physics, California Institute of Technology, Pasadena,
California 91125, USA}

\begin{abstract}
A prominent signature of Majorana bound states is the exotic Josephson effects they produce, the classic example being a fractional Josephson current with $4\pi$ periodicity in the phase difference across the junction. Recent work established that topological insulator edges support a novel `magneto-Josephson effect', whereby a dissipationless current exhibits $4\pi$-periodic dependence also on the relative orientation of the Zeeman fields in the two banks of the junction. Here, we explore the magneto-Josephson effect in junctions based on spin--orbit-coupled quantum wires.  In contrast to the topological insulator case, the periodicities of the magneto-Josephson effect no longer follow from an exact superconductor-magnetism duality of the Hamiltonian.  We employ numerical calculations as well as analytical arguments to identify the domain configurations that display exotic Josephson physics for quantum-wire junctions, and elucidate the characteristic differences with the corresponding setups for topological 
insulators edges.  To provide guidance to experiments, we also estimate the magnitude of the magneto-Josephson effects in realistic parameter regimes, and compare the Majorana-related contribution to the coexisting $2\pi$-periodic effects emerging from non-Majorana states.
\end{abstract}

\maketitle

\section{Introduction}

Given their exotic properties and intriguing promise for topological quantum information processing,\cite{kitaev03,freedman98} Majorana fermions have recently received much attention in the condensed-matter context.\cite{review1,review2,review3} Promising habitats of Majorana fermions include the $\nu=5/2$ fractional quantum Hall state \cite{read00} as well as topological insulator edges \cite{fu08,fu09} or semiconductor quantum wires\cite{lutchyn10,oreg10} proximity coupled to $s$-wave superconductors. Several experiments may have already provided evidence for Majorana bound states in semiconductor quantum wires.\cite{mourik12,das12,churchill13,rokhinson12,lund,harlingen} One of the most direct but also challenging experimental confirmations of the existence of Majorana bound states would be based on the periodicity of the Josephson effect. For junctions of topological superconductors, the Josephson effect is predicted to be $4\pi$ periodic in the phase difference of the order parameter, in sharp contrast 
to the conventional $2\pi$ periodicity\cite{kitaev01,kwon04,fu09} (see also Refs.\ \onlinecite{driss11,jiang11,law,jose12,dom,ojanen13,yokoyama13,houzet13} for more recent works).

Recently, it was noticed that a topological-insulator edge, proximity coupled to an $s$-wave superconductor exhibits an exact superconductivity-magnetism duality.\cite{nilsson08,jiang13} The duality transformation maps the phase of the superconducting order parameter to the direction of the applied magnetic field in the plane perpendicular to the spin--orbit field. As a consequence, the duality predicts a magneto-Josephson effect by which a rotation of the magnetic field across a junction induces a Josephson current even in the absence of a phase gradient.\cite{jiang13,meng12,kotetes13}

Explicitly, proximity-coupled topological-insulator edges are described by the Bogoliubov-de Gennes Hamiltonian\cite{fu08}
\begin{align}
{\mathcal{H}}_{\rm TI}=v\hat{p}\tau^{z}\sigma^{z}  &  -\mu\tau^{z}+\Delta\left(
\cos\phi~\tau^{x}-\sin\phi~\tau^{y}\right) \nonumber\\
&  -b\sigma^{z}+B\left(  \cos\theta~\sigma^{x}-\sin\theta~\sigma^{y}\right)  .
\label{eq:DefH}
\end{align}
Here we have employed the Nambu spinor basis $\Psi^{T}=(\psi_{\uparrow},\psi_{\downarrow},\psi_{\downarrow}^{\dagger},-\psi_{\uparrow}^{\dagger})$
and introduced Pauli matrices $\sigma^{a}$ and $\tau^{a}$ that act in the spin and particle-hole sectors, respectively. The edge-state velocity is given by $v$, $\hat{p}$ is the momentum, and the $\sigma^{z}$-direction represents the spin--orbit-coupling axis. We allow the chemical potential $\mu$,
superconducting pairing $\Delta e^{i\phi}$, longitudinal magnetic field strength $b$, transverse magnetic field strength $B$, and the transverse-field
orientation angle $\theta$ to vary spatially. This Hamiltonian takes the same form upon interchanging the magnetic terms $\left\{  b,B,\theta,\sigma^{a}\right\}  $ with the superconducting terms $\left\{  \mu,\Delta,\phi,\tau^{a}\right\} $. An important aspect of this duality is that it maps the two topologically distinct phases of the model into each other, mapping the `$\Delta$-phase' (occurring for $\Delta^2-b^2 > {\rm max}\{B^2-\mu^2,0\}$) into the `$B$-phase' (occurring for $B^2-\mu^2 > {\rm max}\{\Delta^2 - b^2,0\}$) and vice versa. 

For topological-insulator edges, the duality immediately allows one to derive the periodicity of the magneto-Josephson effect from the known periodicities of the Majorana Josephson effect.\cite{jiang13,meng12,kotetes13} To this end, we consider three-leg junctions \cite{jiang11} with the phase arrangements $B-\Delta-B$ (with a $2\pi$-periodic Majorana Josephson effect) and $\Delta-B-\Delta$ (with a $4\pi$-periodic Majorana Josephson effect). The duality implies that the periodicities are reversed for the magneto-Josephson effect, which is $4\pi$ periodic in the magnetic-field orientation for $B-\Delta-B$ junctions but $2\pi$ periodic for a $\Delta-B-\Delta$ setup. Strictly speaking, the duality also maps charge Josephson currents into spin Josephson currents. At first sight, this may suggest that a change in direction of the magnetic field across a junction only drives a spin Josephson current. However, it was shown in Ref.\ \onlinecite{jiang13} that as a result of the spin-momentum locking, there is also a 
conventional (and experimentally more accessible) charge current across the junction in addition to the spin current. 

While the magneto-Josephson effect has been studied in some detail for topological insulator edges,\cite{jiang13,meng12,kotetes13} much less is known about it for junctions based on semiconductor quantum wires. There are several reasons why this poses an interesting problem. Many of the ongoing searches for Majorana fermions are based on quantum-wire based structures. There are also several distinct differences between topological superconducting phases based on proximity-coupled topological insulators and semiconductor quantum wires. First, the kinetic energy of the quantum-wire Hamiltonian explicitly violates the duality, making the duality only of suggestive value for the quantum-wire situation. Second, the two topologically distinct phases of the topological insulator effectively trade places in the quantum wire. For instance, a $4\pi$-periodic Majorana Josephson effect occurs in the $ \Delta - B - \Delta$ arrangement in topological insulators, but in the $B - \Delta - B$ arrangement in quantum wires.  

This motivates us to explore the magneto-Josephson effects in semiconductor quantum wires in more detail in this paper. In Sec.\ \ref{sec:numerical}, we present numerical results based on a recursive scattering-matrix approach and establish the periodicities of the magneto-Josephson effects. In Sec.\ \ref{sec:limits}, we provide further insight into the periodicities by analytical arguments and the analysis of limiting cases. Finally, Sec.\ \ref{sec:estimates} is concerned with numerical estimates of the magnitude of the effect and Sec.\ \ref{sec:conclusions} collects our conclusions. 

\begin{figure}[tp]
\centering
\includegraphics[width=.35\textwidth]{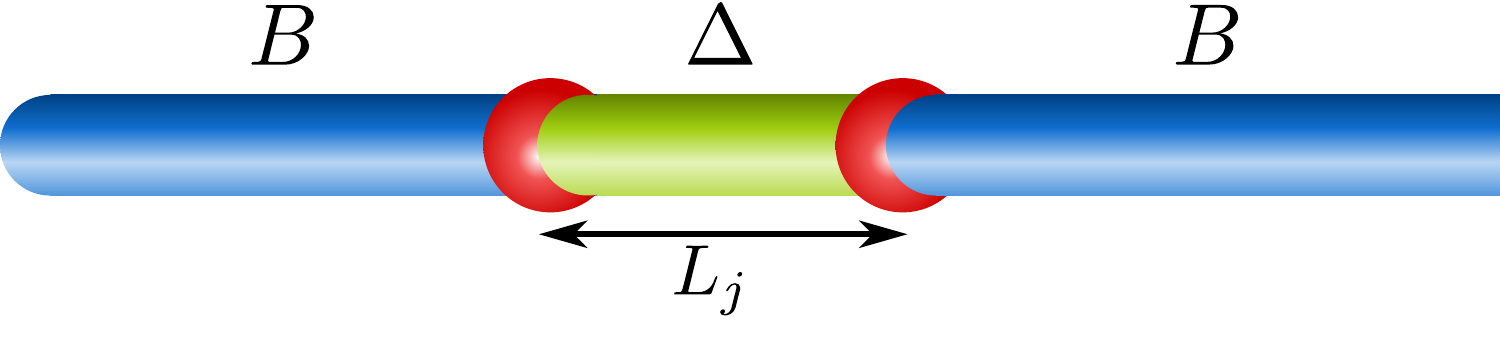}\\
\includegraphics[width=.23\textwidth]{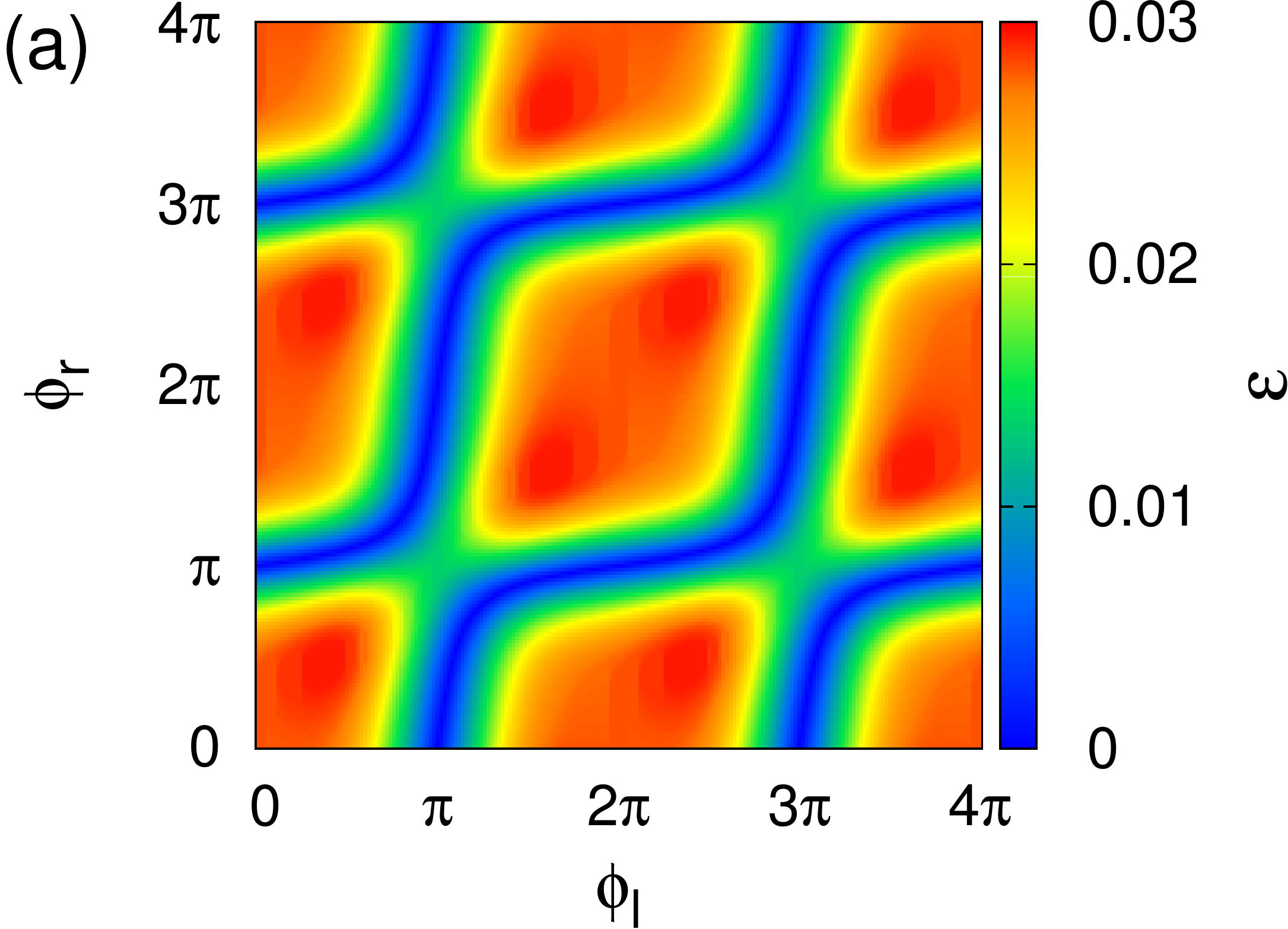}
 \includegraphics[width=.23\textwidth]{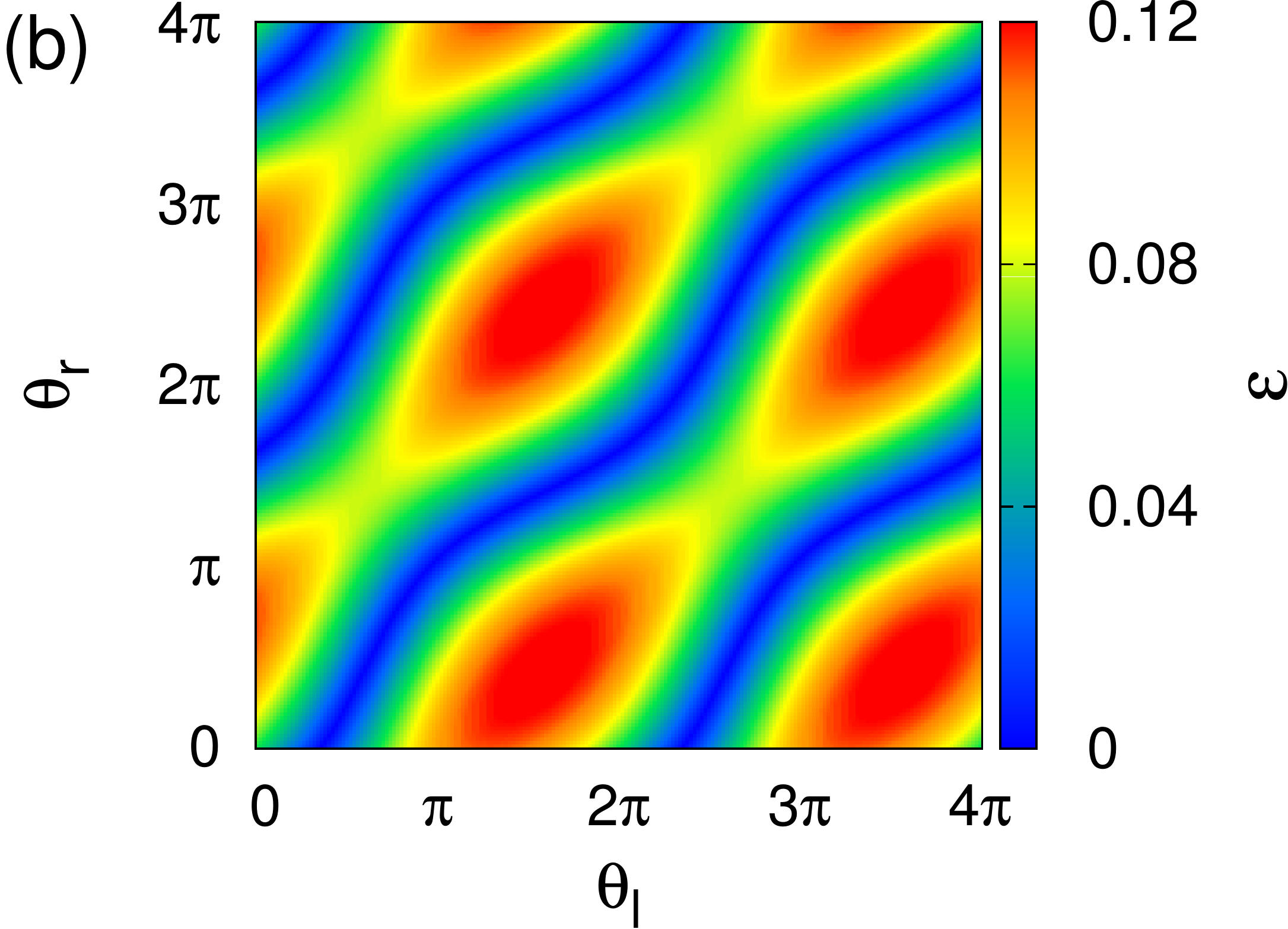} \\
\includegraphics[width=.23\textwidth]{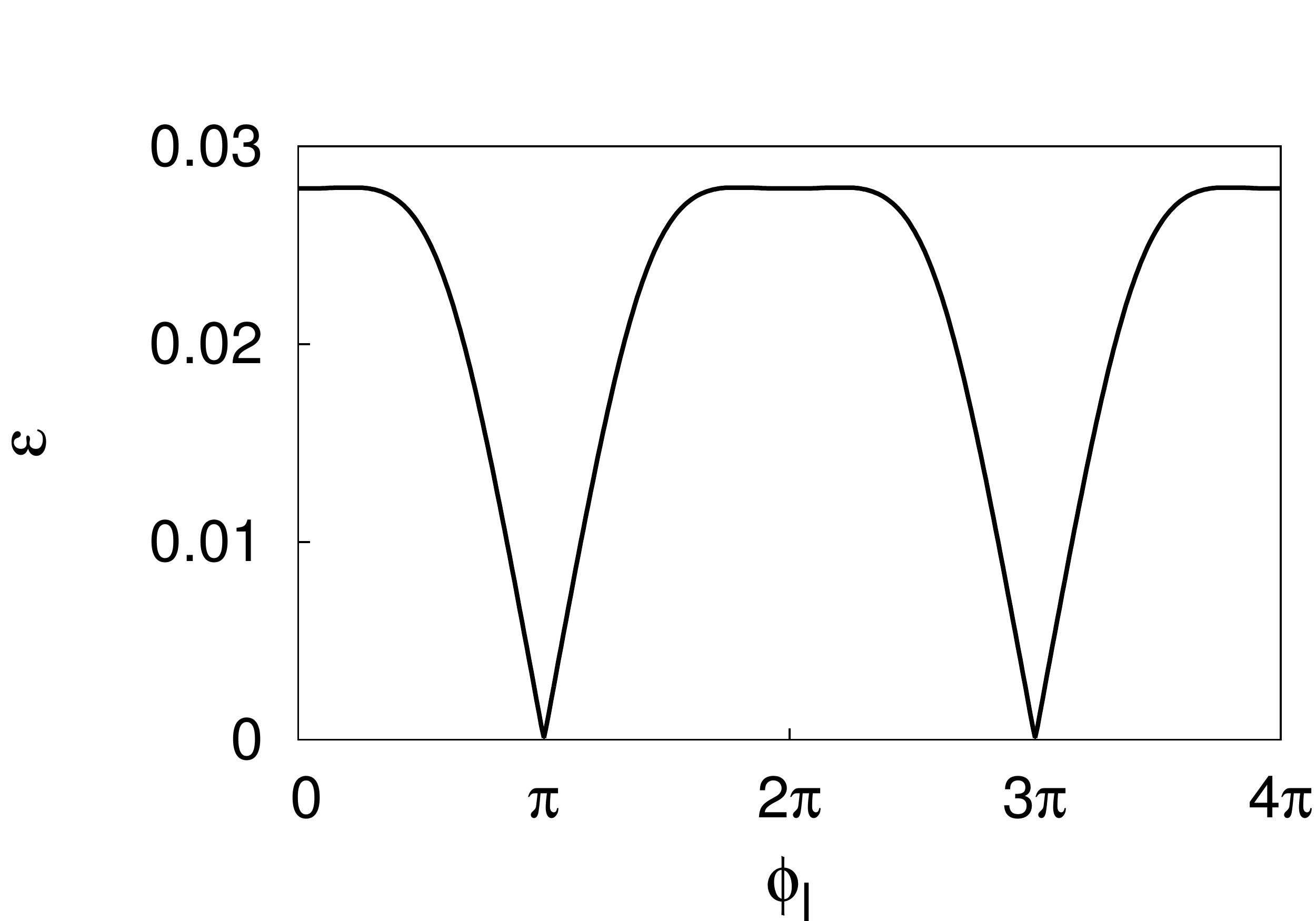}
 \includegraphics[width=.23\textwidth]{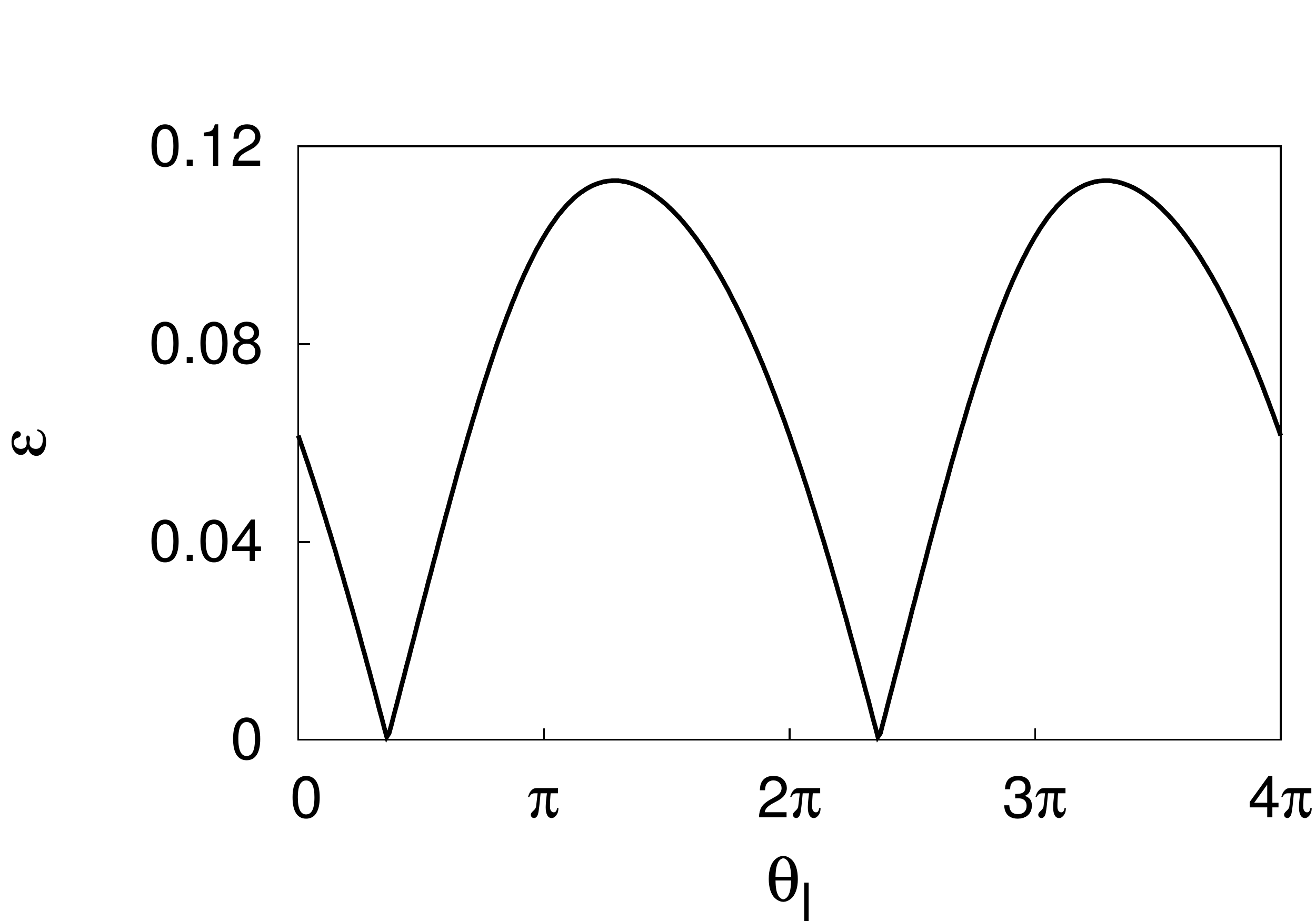} 
\caption{Upper panels: Color scale plots of the low-energy Bogoliubov-de Gennes spectra (only $\epsilon>0$) of a $B-\Delta-B$ junction vs.\ (a) the superconducting phases $\phi_l$ and $\phi_r$ and (b) the magnetic-field directions $\theta_l$ and $\theta_r$. Lower panels:  Corresponding line cuts along $\phi_r=0$ and $\theta_r=0$, respectively. Both line cuts exhibit cusps at zero energy, reflecting protected zero-energy crossings and thus, the Josephson current is $4\pi$ periodic in both $\phi$ and $\theta$. The parameters for the three segments are $\Delta_{l/r}=1.6$, $\Delta_m=2$, $B_{l/r}=2$, $B_m=0.9$, $\mu_{l/m/r}=0$, and $L_j=2$ (length of the junction). We also set $\theta_l=\theta_r=\pi/2$, $\theta_m=0$ in (a) 
and $\phi_l=\phi_r=\pi/2$, $\phi_m=0$ in (b). Note that the parameters of the three segments are labeled by subscripts $l$, $m$, and $r$.}
\label{fig:TST}
\end{figure}

\section{Numerical results}
\label{sec:numerical}

We now turn to semiconductor quantum wires proximity coupled to $s$-wave superconductors. The Hamiltonian for a clean, single-channel semiconductor quantum wire (QW) in the presence of a Zeeman field $B$, Rashba spin--orbit coupling $u$, and induced superconductivity $\Delta$ is\cite{lutchyn10,oreg10}
\begin{align}
\begin{split}
 {\mathcal{H}}_{\rm QW}=&\left(\frac{\hat{p}^2}{2m}-\mu\right)\tau_z+u\hat{p}\sigma_z\tau_z + B\left(e^{i\theta}\sigma_++e^{-i\theta}\sigma_{-}\right)\\
 & + \Delta\left(e^{i\phi}\tau_++e^{-i\phi}\tau_{-}\right)\label{hamil_QW}
\end{split}
\end{align}
Other than dropping the longitudinal magnetic field term for lack of relevance in the following, this Hamiltonian differs from that of the topological insulator edge in Eq.\ (\ref{eq:DefH}) by the kinetic term $\hat{p}^2/2m$. This term explicitly breaks the duality present for the topological insulator edge and is responsible for key differences between the topological insulator edge and the quantum wire. Most importantly, the phases are in some sense effectively reversed in the two systems. Explicitly, in quantum wires, the topological (or $B$) phase occurs for $B^2>\Delta^2 + \mu^2$, while the nontopological (or $\Delta$) phase requires $B^2 < \Delta^2 + \mu^2$. In the quantum wire model, the identification of topological and nontopological phases is unique since the $\Delta$-phase is continuously connected to the vacuum. The corresponding identification is less defined for the topological insulator as the model does not connect naturally to the vacuum due to the linear spectrum. Indeed, the duality of the 
model maps the two phases into each other, suggesting that they are topologically distinct but cannot be labeled as topological and nontopological. However, if we take the presence or absence of the fractional ($4\pi$-periodic) Majorana Josephson effect as the defining feature of a topological superconducting phase, we would crudely label the $\Delta$ phase as topological and the $B$ phase as nontopological, which just reverses the assignments for the quantum wire model. 
   
\begin{figure}[tp]
\centering
\includegraphics[width=.35\textwidth]{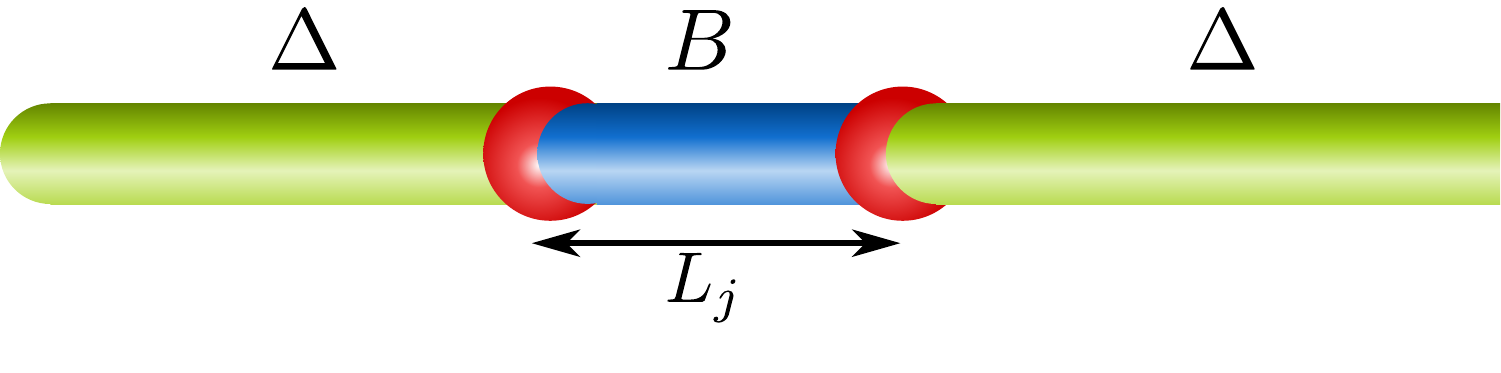}\\
\includegraphics[width=.23\textwidth]{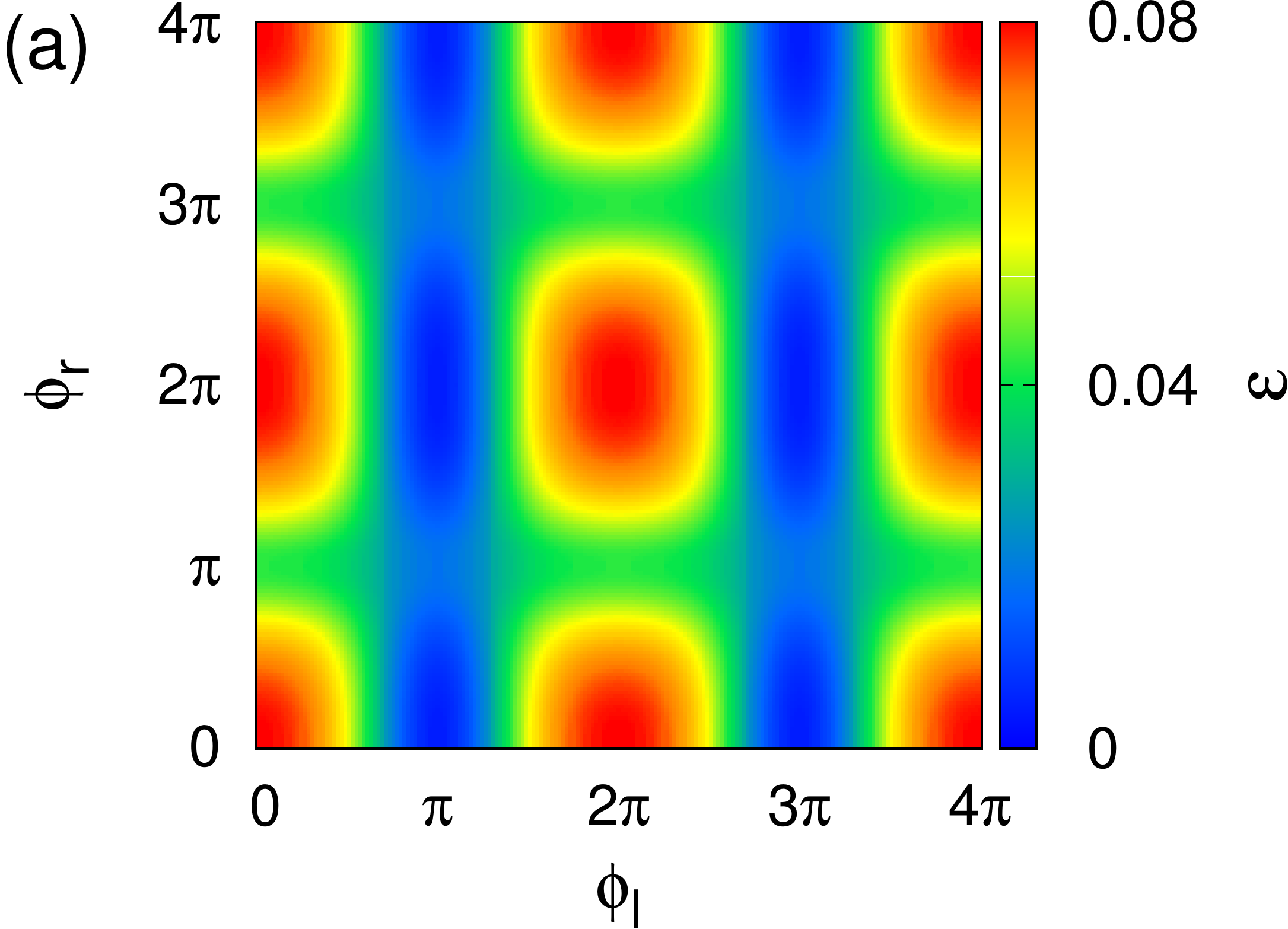}
\includegraphics[width=.23\textwidth]{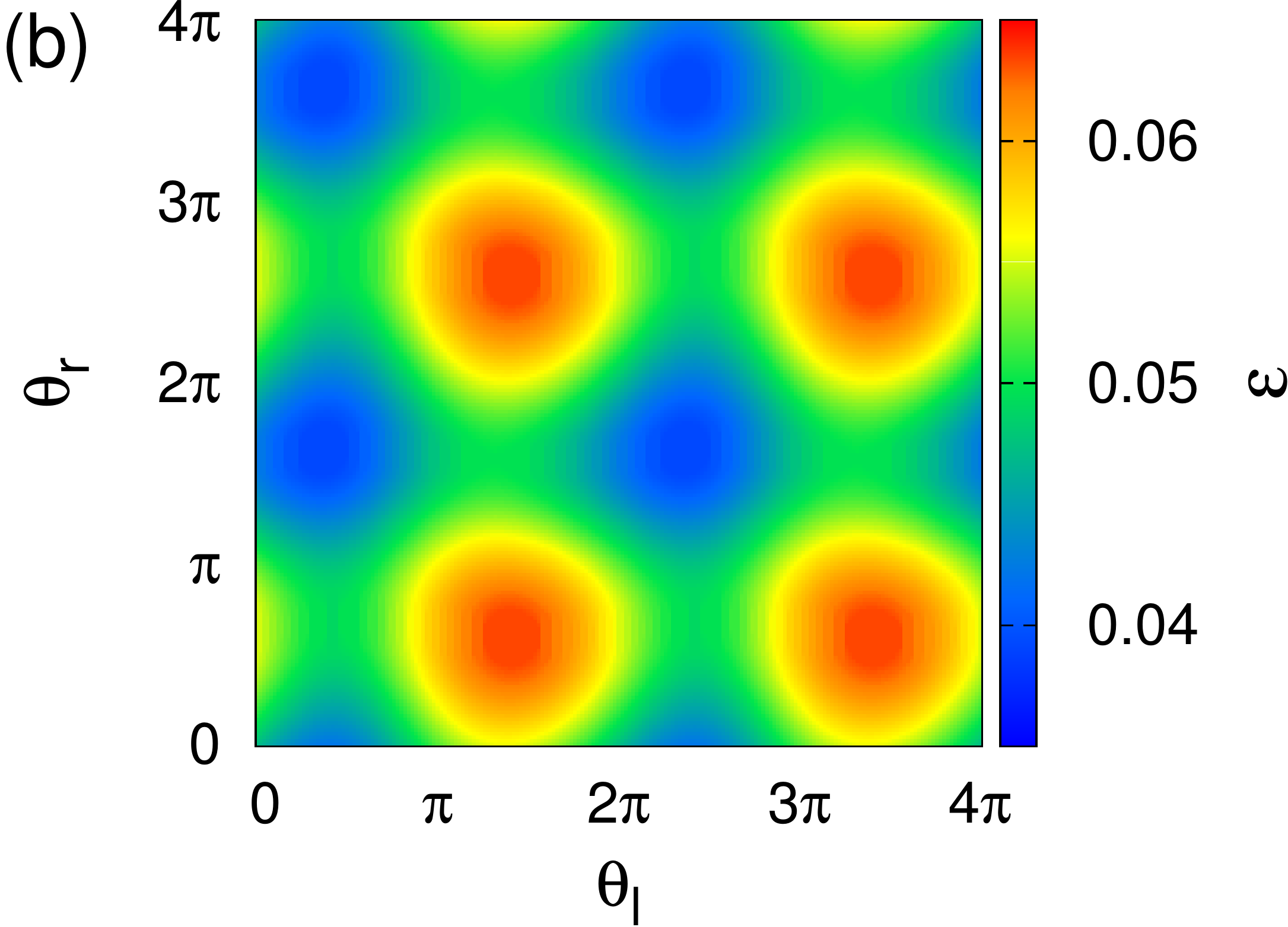} \\
\includegraphics[width=.23\textwidth]{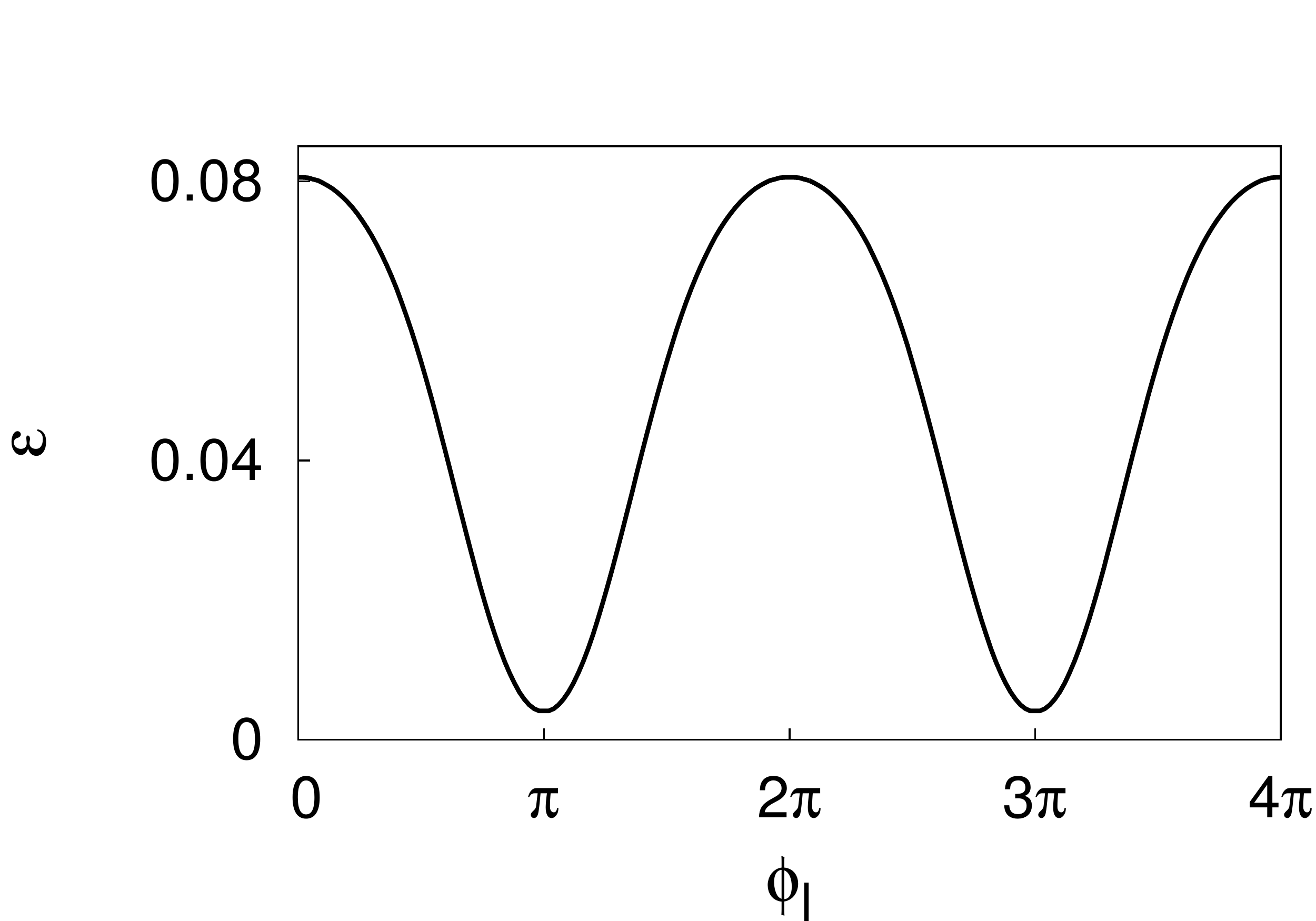}
\includegraphics[width=.23\textwidth]{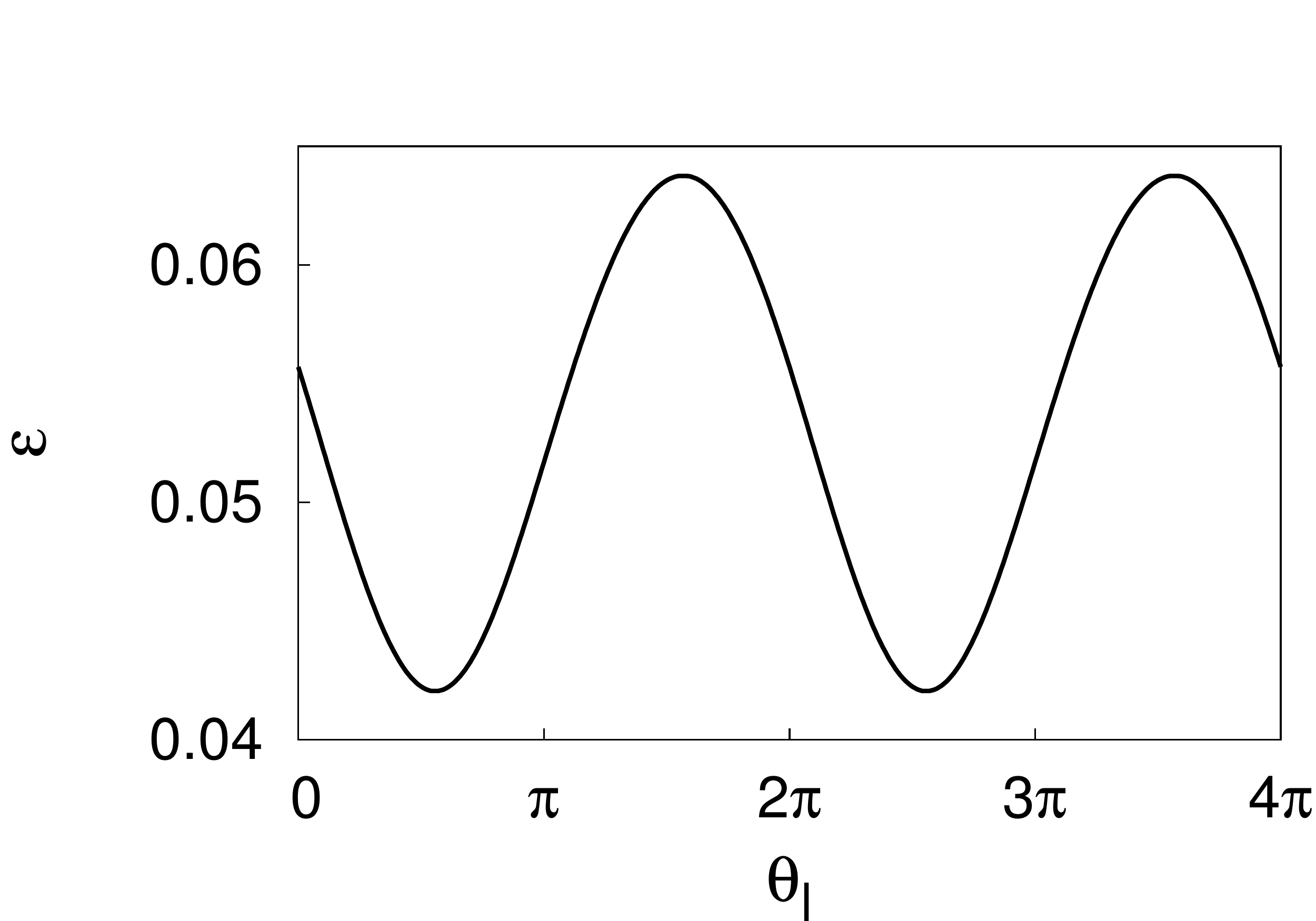} 
\caption{Upper panels: Color scale plots of the low-energy Bogoliubov-de Gennes spectra (only $\epsilon>0$) of a $\Delta-B-\Delta$ junction as a function of (a) superconducting phases and (b) magnetic field directions. Lower panels:  Corresponding line cuts along $\phi_r=0$ and $\theta_r=0$, respectively. There are no zero-energy crossings and the current is $2\pi$ periodic. The parameters for the three segments are $\Delta_{l/r}=2$, $\Delta_m=1.6$, $B_{l/r}=0.9$, $B_m=2$, $\mu_{l/m/r}=0$, and $L_j=5$. We also set $\theta_l=\theta_r=\pi/2$, $\theta_m=0$ in (a) and $\phi_l=\phi_r=\pi/2$, $\phi_m=0$ in (b).}
\label{fig:STS}
\end{figure}

We consider quantum-wire junctions consisting of three segments, with phase arrangements $B - \Delta - B$ and $\Delta - B - \Delta$. It is well established that the periodicity in the superconducting phase difference across the junction is $4\pi$ in the $B - \Delta - B$ arrangement, but $2\pi$ for the $\Delta - B - \Delta$ setup.\cite{lutchyn10,oreg10} These periodicities are reproduced in our numerical calculations of the low-energy Bogoliubov-de Gennes spectra shown in Figs.\ \ref{fig:TST}(a) and \ref{fig:STS}(a). Here, we restrict ourselves to $\mu=0$ for simplicity. The calculations are based on a scattering-matrix approach which has been employed previously in the context of topological superconducting phases and Majorana fermions in quantum wires (see, e.g., Refs.\ \onlinecite{brouwer11} and \onlinecite{pientka12}). In short, it is based on concatenating small slices of quantum wire to obtain the scattering matrix $S(\epsilon)$ of the entire wire. The spectrum can then be determined by solving the 
equation ${\rm det}[1-S(\epsilon)]=0$. A more detailed description of the method can be found in Ref.\ \onlinecite{brouwer11}.

Fig.\ \ref{fig:TST}(a) shows the low-energy spectrum of a $B-\Delta-B$ junction as a function of the phases of the superconducting order parameters of the outer segments. The left and right segments are chosen much longer than the coherence length so that the Majorana bound states at the outer ends do not couple to the Majoranas at the junction and can be safely ignored. The low-energy spectrum shows protected zero-energy crossings, which makes the current $4\pi$ periodic as a function of $\phi$. The corresponding spectrum of a $\Delta-B-\Delta$ junction is shown in Fig.~\ref{fig:STS}(a). In contrast to the $B-\Delta-B$ case, the current is always $2\pi$ periodic. For both types of junctions, there are two Majorana bound states at the interfaces between the $B$ and $\Delta$ regions. However, in the latter case the hybridization of the Majoranas does not generate a protected crossing at zero.

Representative results for the dependence of the low-energy Bogoliubov-de Gennes spectra of the junctions on the directions $\theta$ of the magnetic fields are shown in Figs.\ \ref{fig:TST}(b) and \ref{fig:STS}(b). We find that also the dependence on $\theta$ is $4\pi$ periodic for $B - \Delta - B$ junctions and $2\pi$ periodic for $\Delta - B - \Delta$ junctions, with the spectra exhibiting protected zero-energy crossings in the first case, but not in the second. We summarize the periodicities for the two types of junctions in quantum wires in Table \ref{tab:periodic_table}. Remarkably, the magneto-Josephson effect has the same periodicity for the quantum wire and the topological-insulator edge. This is in stark contrast with the ordinary Josephson current which has different periodicities in the two models reflecting the reversed roles of topological and nontopological phases. 

\begin{table}[tb]
\caption{Periodicities of the Josephson energy as a function of the phase difference of the superconducting gap and the relative magnetic-field orientation $\theta$. We list results for $B-\Delta-B$ and $\Delta-B-\Delta$ junctions realized in quantum wires (QW) and topological insulator edges (TI). The latter results are taken from Ref.~\onlinecite{jiang13}.\\}\label{tab:periodic_table}

\begin{ruledtabular}
 \begin{tabular}{ccccc}
& \multicolumn{2}{c}{$B-\Delta-B$}  & \multicolumn{2}{c}{$\Delta-B-\Delta$}\\ \cline{2-3}\cline{4-5}
 & $\phi$ & $\theta$ & $\phi$ & $\theta$\\
\hline
periodicity for QW & $4\pi$& $4\pi$ & $2\pi$ & $2\pi$\\
periodicity for TI edge & $2\pi$ & $4\pi$ & $4\pi$ & $2\pi$
\end{tabular}
\end{ruledtabular}
\end{table}

\section{Limiting cases and analytical considerations}
\label{sec:limits}

To gain more insight into the periodicities of the Josephson effects summarized in Table \ref{tab:periodic_table} and their relations, we now combine analytical arguments and an analysis of limiting cases. First, we use analytical arguments to derive the Josephson periodicities for quantum wires which are based on the well-established result that the dependence on the superconducting phase is $4\pi$ periodic for a $B - \Delta - B$ junction (fractional Josephson effect). This complements the arguments based on the magnetism-superconductivity duality for the topological insulator edge. 

To gain a better understanding of the similarities of and differences between the topological insulator and quantum wire cases, we then study the limit of large spin--orbit coupling $\epsilon_{\rm SO} = mu^2$ for the quantum wire model (or equivalently large mass $m$), i.e.\ $\epsilon_{\rm SO}\gg \Delta\gg |B-\Delta|$. In this limit, there are strong similiarities between the low-energy spectra of the topological-insulator edge and the quantum wire.  

\subsection{Analytical argument}
\label{analytical}

In this section, we derive the periodicities in both $\phi$ and $\theta$ for the quantum-wire case by analytical arguments. Our arguments assume the well-established fractional Josephson effect (i.e.\ a $4\pi$-periodic $\phi$ dependence) for a $B - \Delta - B$ junction and reproduce the numerically obtained periodicities summarized in Table \ref{tab:periodic_table}.  To this end, it suffices to derive the number of protected zero-energy crossings
considering particular limiting cases of the two types of junctions. By adiabatic continuity, these periodicties must then hold for junctions
of the same kind with arbitrary parameters. 

Consider first a quantum-wire junction in the $\Delta-B-\Delta$ configuration. The $\Delta$ phase of the quantum wire is adiabatically connected to the vacuum by making the chemical potential large and negative. At the same time, the $B$ phase is adiabatically connected to a spinless $p$-wave superconductor by taking the limit of large Zeeman field $B$.\cite{alicea11} Consequently, a $\Delta-B-\Delta$ junction can be adiabatically deformed into an essentially finite segment of a $p$-wave superconducting wire with hard-wall boundary conditions. In this limit, the two Majorana bound states localized at the domain walls hybridize and split by some finite energy. Clearly, the two Majorana bound states will penetrate only very little into the $\Delta$ sections of the junctions, and consequently, they will be only weakly dependent on $\phi$ and $\theta$ as long as $|B|,|\Delta| \ll |\mu|$, where $\mu<0$ is the chemical potential in the outer $\Delta$ segments of the wire.
Thus, while there will be a variation of the energy splitting with $\phi$ and $\theta$, it will be small compared to the magnitude of the splitting itself. 
Thus, there are no zero-energy crossings in this case and the Josephson current is $2\pi$ periodic both in $\theta$ and $\phi$.
These considerations are only valid for the quantum wire because the decay length of the Majorana states into the insulating segments on the outside is controlled by $|\mu|$. In a TI-edge junction the gap in the $\Delta$ segments is controlled by the pairing strength $\Delta$ and not by $\mu$. Therefore, the effect of $\Delta$ in TI edges is never perturbative and the above argument does not hold for $\phi$.

We now turn to the $B-\Delta-B$ junction for which the Majorana energy is $4\pi$ periodic both in $\phi$ and $\theta$. The $4\pi$ periodicity as a
function of $\phi$ represents the well-known fractional Josephson effect.\cite{kitaev01,kwon04,fu09} In the remainder of this section, we demonstrate that the parities of the number of protected zero-energy crossings of the Majorana energy dispersion
as a function of $\phi$ and $\theta$ are equal for a $B-\Delta-B$ quantum wire junction. The basic observation is that we can again consider the limit in which the middle $\Delta$ section has a large and negative $\mu$. In this insulating limit, the gap does not close when we take $B$ and $\Delta$ equal to zero. In effect, we can thus replace the $B - \Delta -B$ junction by a $B-I-B$ junction, where the middle section is a conventional insulator ($I$). 

We start by considering a $B-I$ interface between a $B$ dominated phase with $\phi,\theta=0$ and a normal insulator with $B,\Delta=0$ and
$\mu<0$. This interface harbors one zero-energy Majorana bound state with wavefunction $\psi$. We can tune the left region to the phase $\phi$ and the
angle $\theta$ by performing the unitary transformation $U(\phi,\theta)=\exp(i\phi\tau_z/2+i\theta\sigma_z/2)$ on the Majorana wavefunction, i.e.,
$\psi(\phi,\theta)=U(\phi,\theta)\psi$. It is crucial for our argument that we can effect the variation of $\phi$ and $\theta$ in the left region by a {\em global} transformation $U(\phi,\theta)$, which is possible because the rotation of $B$ and $\Delta$ does not affect the normal insulator on the right. We note that $U(2\pi,0)=U(0,2\pi)=-1$, which guarantees that the Majorana wavefunction evolves to the same final state, when either $\phi$ or
$\theta$ advance by $2\pi$.

We now consider weak coupling of two such interfaces in a $B-I-B$ junction. This coupling leads to a symmetric splitting of the two Majorana states about zero energy. When the coupling across the junction is sufficiently weak, we can obtain this subgap spectrum emerging from the Majorana modes localized at the junction accurately from first-order perturbation theory. Starting at $\phi=\theta=0$ and tuning either $\phi$ or $\theta$ to $2\pi$ the initial wavefunction evolves to the same final state. Consequently, the initial and final subgap-energy spectra emerging from the hybridized Majorana modes will be identical for both processes. We know from the fractional Josephson effect that the positive-energy excitation at $\phi=0$ becomes negative (and vice versa) when $\phi$ advances by $2\pi$, and hence the associated Bogoliubov-de Gennes eigenenergy must cross zero energy an odd number of times in the process. Given that $U(2\pi,0)=U(0,2\pi)=-1$, this immediately implies that the positive- and negative-energy 
excitations also exhibit an odd number of zero-energy crossings when tuning $\theta$ from 0 to $2\pi$ instead, which proves the $4\pi$ periodicity as a function of $\theta$.

It is worth noting that this argument fails for the TI edge model (\ref{eq:DefH}), as it should according to Table \ref{tab:periodic_table}. The reason is that irrespective of $\mu$, the corresponding spectrum is never gapped when setting $B,\Delta=0$.

\subsection{Strong spin--orbit coupling ($\epsilon_{\rm SO}\gg B> \Delta$)}
\label{strongSO}

The arguments in the previous subsection explain the periodicities of the magneto-Josephson effects for semiconductor quantum wires. When combined with the duality arguments for topological insulator edges, this explains the full set of periodicities collected in Table \ref{tab:periodic_table}.
How the periodicities of these two systems are related, however, remains an open question. This is particularly interesting in the limit $\epsilon_{\rm SO}\gg B> \Delta$, when the low-energy bulk spectrum of the quantum wire is nearly identical to the spectrum of a topological insulator edge.

\begin{figure}[tp]
\centering
\includegraphics[width=.48\textwidth]{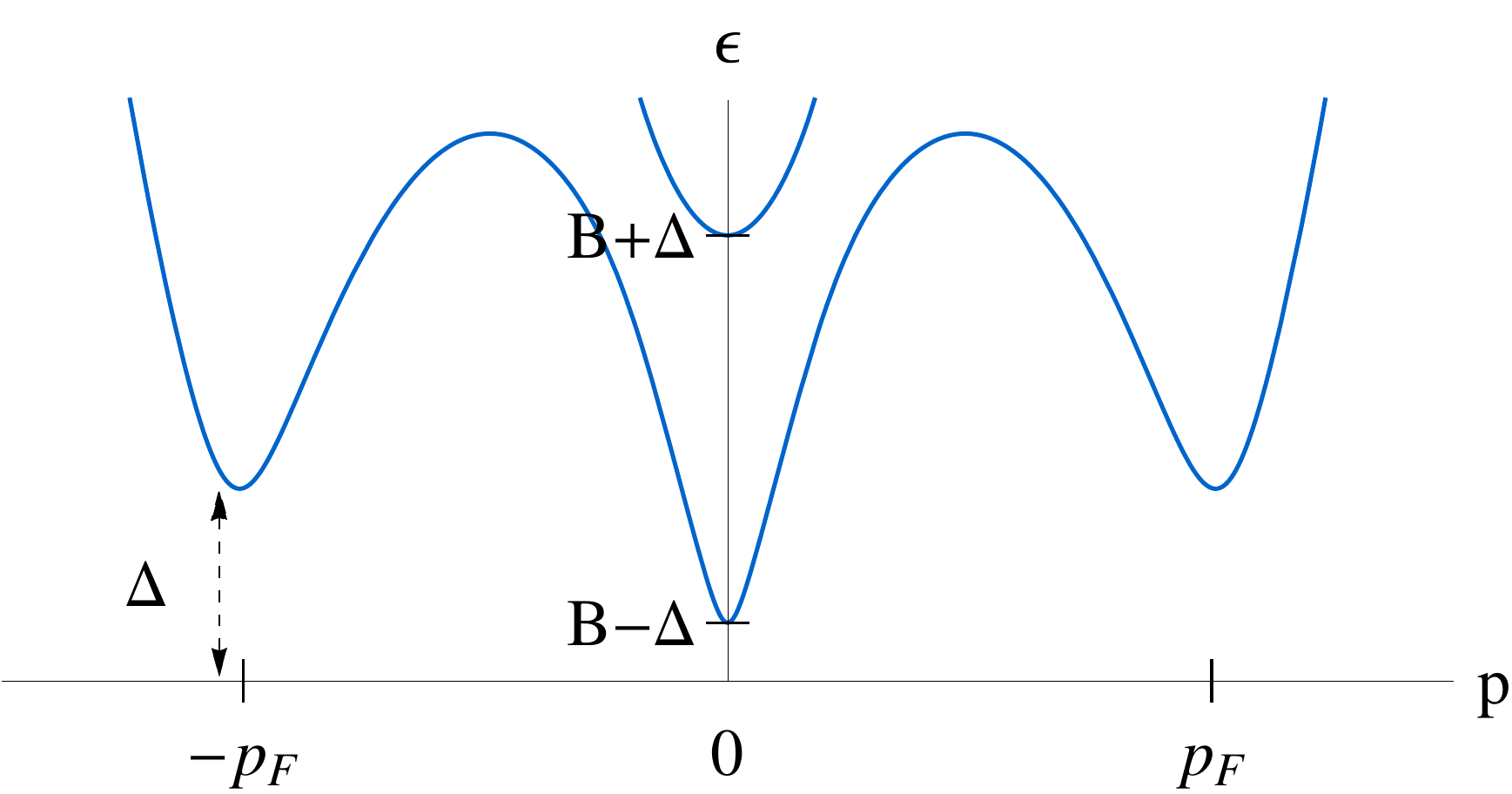}
\caption{Bulk spectrum of the quantum-wire Hamiltonian (\ref{hamil_QW}) in the limit of strong spin--orbit coupling.}
\label{fig:bulk}
\end{figure}

When the spin--orbit energy is much larger than the Zeeman energy, the bulk spectrum of Eq.~(\ref{hamil_QW}) depicted in Fig.~\ref{fig:bulk} has three minima located at $p=0$ and $p=\pm p_F$, where $p_F=2mu$ when $\mu=0$. Since $p_F$ is large in the limit of strong spin--orbit coupling, the subspaces at $p=0$ and at $p=\pm p_F$ effectively decouple for sufficiently smooth domain walls and the low-energy spectrum can be understood as arising from a superposition of two subspectra.\footnote{Note that the superconducting pairing couples the states near $+p_F$ with those near $-p_F$ so that these momenta cannot be considered separately.} Near $p=0$, the Hamiltonian (\ref{hamil_QW}) can be linearized and reduces to the Hamiltonian of the topological insulator edge (i.e., Eq.\ (\ref{eq:DefH}) with $\mu = b = 0$). Near $p=\pm p_F$, the Hamiltonian can be linearized, as well, and reduces to that of a spinless $p$-wave superconductor (cf.\ App.\ \ref{appendix}). This describes a topological superconductor by itself. 
Since the  topologically distinct phases are labelled by a ${\bf Z}_2$ index, this provides an explanation for the effective reversal of phases between the quantum-wire and the topological-insulator Hamiltonian.

\begin{figure}[tb]
 \begin{center}
 \includegraphics[width=.35\textwidth]{junc_BDB1.pdf}\\
\includegraphics[width=.45\textwidth]{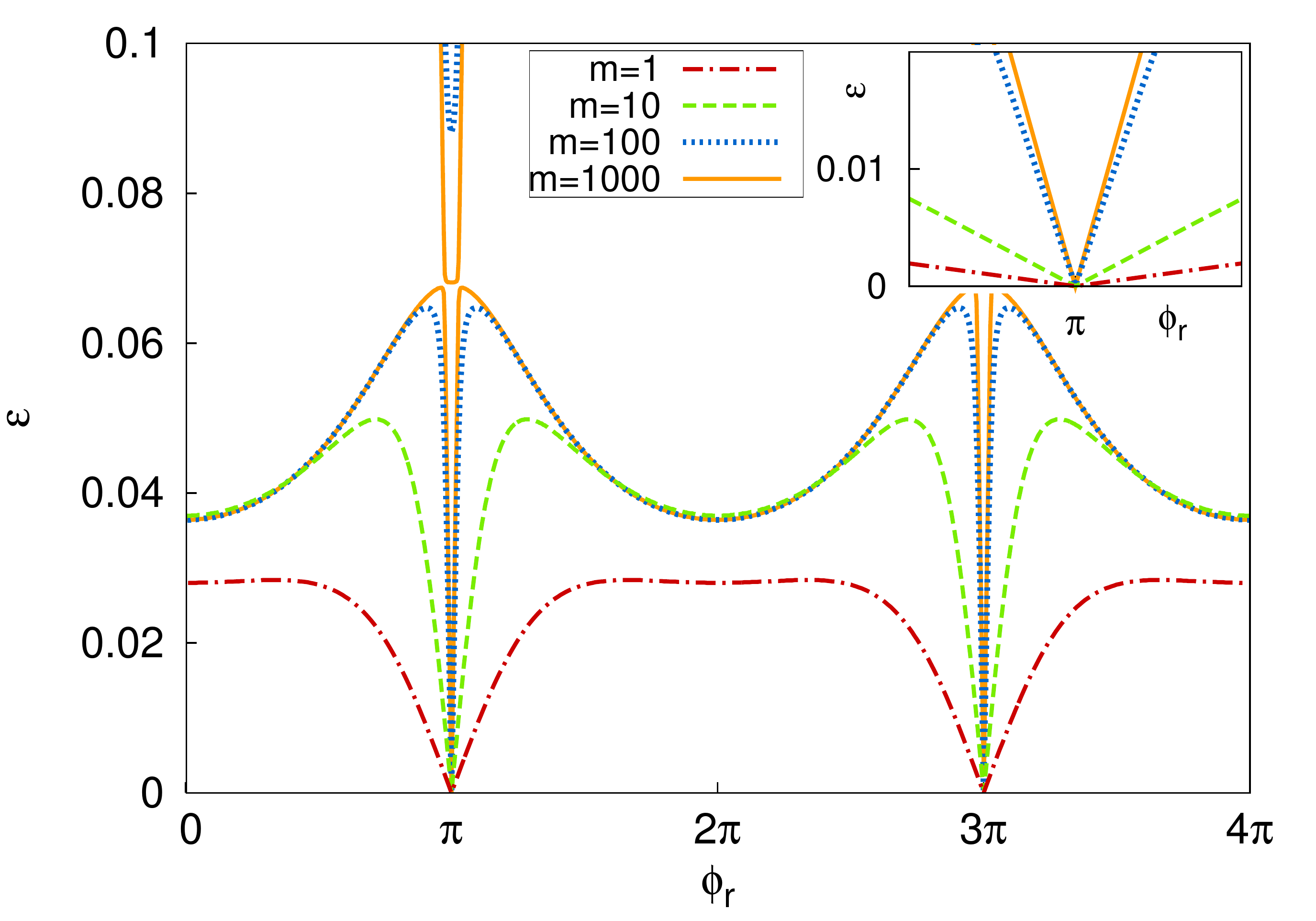}
\end{center}
\caption{Low-energy spectrum of a $B-\Delta-B$ junction as a function of $\phi_r$ for different masses $m$ (and hence spin--orbit energies $\epsilon_{\rm SO}=mu^2$). For large $\epsilon_{\rm SO}$, the dispersion of the hybridized Majorana bound states becomes $2\pi$ periodic while additional Andreev bound states cross zero energy at $\phi_r =\pi,3\pi$. The avoided crossings between the Andreev and Majorana bound state excitations vanish in the limit $\epsilon_{\rm SO}\gg B$. The inset shows that the zero-energy crossing at $\phi=\pi$ persists for all values of $m$. Parameters: $B_{l/r}=\Delta_m=2$, $\Delta_{l/r}=B_m=1$, $u=1$, $\mu_{l/m/r}=0$, $L_j=2$, and $\theta_{l/m/r}=\phi_{l/m}=0$.
}\label{fig:TST_TI-limit}
\end{figure}

The high-momentum subspace near $p=\pm p_F$ has a gap of size $\Delta$. In contrast, the low-momentum subspace near $p=0$ has a gap equal to $|B-\Delta|$, which is controlled by the competition of Zeeman and pairing energies and which is much smaller when the system is close to the topological phase transition, $|B-\Delta|\ll\Delta$. Zeroes of the gap in the low-momentum subspace trigger the topological phase transition and thus, the Majorana bound states, localized at domain walls between $B$- and $\Delta$-dominated regions, predominantly reside in this subspace. Consequently, in this subspace the relevant subgap spectrum of a short junction is determined by the hybridization of the Majorana bound states and the periodicities as a function of $\phi$ and $\theta$ are those for the topological-insulator edge. This seems consistent with Table \ref{tab:periodic_table} for the dependences on $\theta$, but not for those on $\phi$.

To understand the full set of periodicities in Table \ref{tab:periodic_table}, we thus need to also consider the high-momentum subspace at $\pm p_F$. In this subspace where the Hamiltonian reduces to that of a spinless $p$-wave superconductor, the effective spin--orbit field is large and hence, the magnetic field is only a small perturbation. The corresponding spectrum should thus depend only weakly on $\theta$. At the same time, variations in $\phi$ can result in a considerable Josephson current. In fact, as the high-momentum subspace by itself constitutes a model of a topological superconductor, the $\phi$ dependence of the corresponding Bogoliubov-de Gennes spectrum exhibits protected zero-energy crossings.

\begin{figure}[tb]
 \begin{center}
 \includegraphics[width=.35\textwidth]{junc_DBD1.pdf}\\
\includegraphics[width=.45\textwidth]{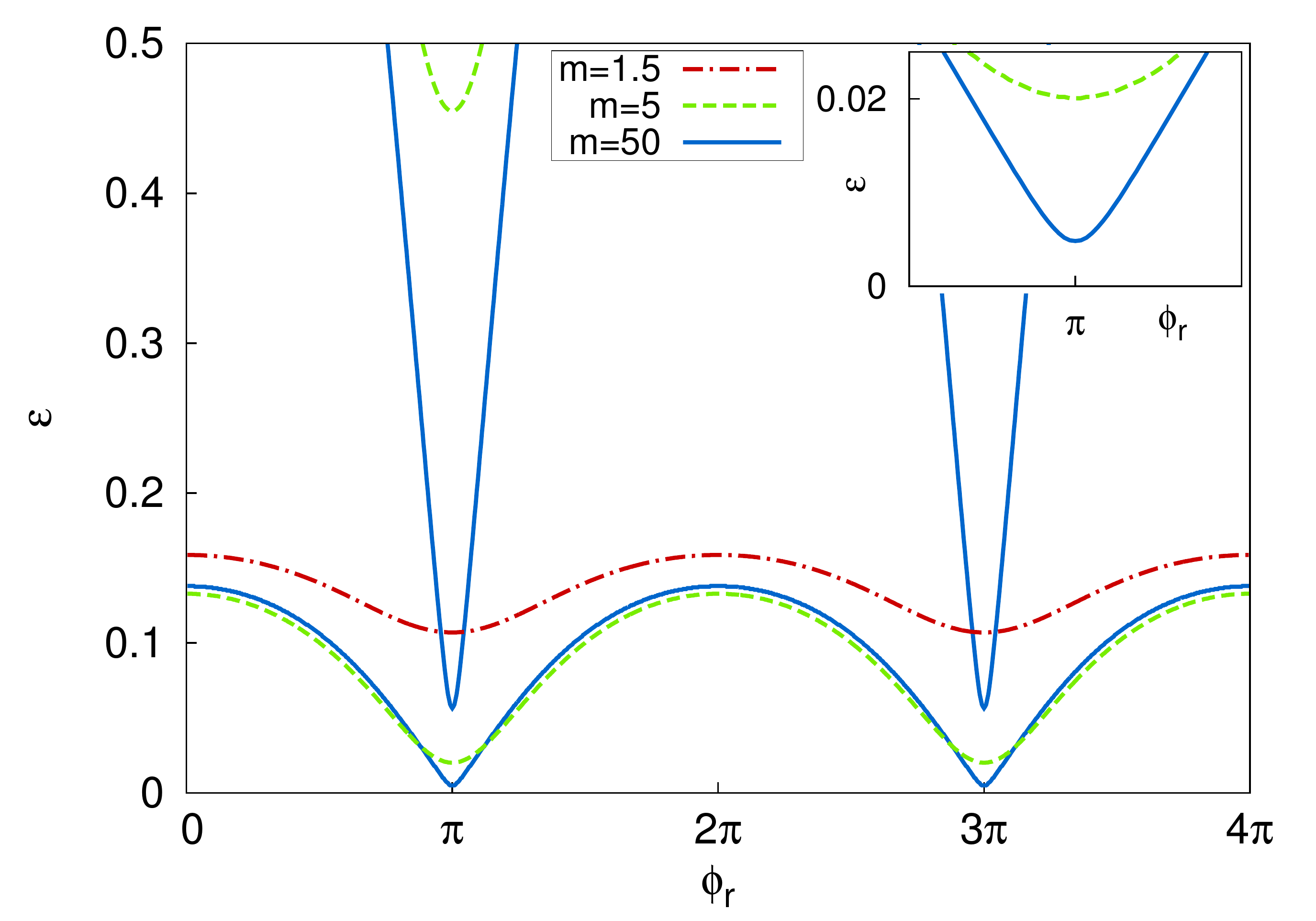}
\end{center}
\caption{Low-energy spectrum of a $\Delta-B-\Delta$ junction as a function of $\phi_r$ for different masses $m$ (and hence spin--orbit energies $\epsilon_{\rm SO}=mu^2$). As indicated by Table \ref{tab:periodic_table}, the spectrum approaches $4\pi$ periodicity for $\epsilon_{\rm SO}\rightarrow\infty$. For strong spin--orbit coupling, Andreev bound states appear. Their energy has a phase dependence $\epsilon_A\sim\Delta\cos(\phi/2)$. Inset: Close-up near $\phi=\pi$ showing the avoided crossing at $\epsilon=0$.  Parameters: $B_{l/r}=\Delta_m=1$, $\Delta_{l/r}=B_m=2$, $u=1$, $\mu_{l/m/r}=0$, $L_j=2$, and $\theta_{l/m/r}=\phi_{l/m}=0$.}
\label{fig:STS_TI-limit}
\end{figure}

We now use these insights to understand the similarities and differences of the Josephson periodicities in topological insulators and quantum wires in more detail. First consider a quantum wire in a $B-\Delta-B$ configuration. Such junctions exhibit a $4\pi$-periodic Josephson current in the superconducting phase, with a protected zero-energy crossing of the Bogoliubov-de Gennes spectra. This contrasts with the $2\pi$ periodicity for the same junction made of topological-insulator edges. To understand this difference in periodicity, Fig.\ \ref{fig:TST_TI-limit} shows how the low-energy spectrum changes with increasing spin--orbit energy. As expected based on the general arguments above, the spectrum develops two distinct types of subgap states as the spin--orbit energy increases, $\epsilon_{\rm SO}\gg B$ (seen most clearly in the traces for $m=1000$ in Fig.\ \ref{fig:TST_TI-limit}). The first type of state has an approximately sinusoidal $\phi$ dependence, an offset from zero energy, and $2\pi$ periodicity. 
This state can be identified with the hybridized Majorana bound states in the low-momentum subspace. The second type of state crosses zero energy at $\phi=\pi,3\pi$ with a $\pm E_0\cos(\phi/2)$ dispersion, where $E_0$ is of the order of $\Delta$. This excitation corresponds to an Andreev bound state at $p=\pm p_F$. As seen in Fig.\ \ref{fig:TST_TI-limit}, there is an avoided crossing between these states which disappears as the spin--orbit energy and, with it, the momentum mismatch diverge. 

This now allows one to understand the periodicities of Table \ref{tab:periodic_table} for the case of $B - \Delta - B$ junctions. In the quantum wire, only the low-momentum subspace has an interesting $\theta$ dependence. Thus, the $\theta$ dependence remains the same between quantum wires and topological-insulator edges. At the same time, both subspaces contribute to the dependence on $\phi$. Indeed, the above considerations show that the protected zero-energy crossing in the quantum-wire spectrum is associated with states which converge entirely on the high-momentum subspace as the spin--orbit energy increases. These states do not exist for the topological-insulator edge whose $\phi$ dependence is thus $2\pi$ periodic. 

In a $\Delta-B-\Delta$ junction, the change of periodicities is opposite. While the quantum wire is $2\pi$ periodic in $\phi$, the topological-insulator edge is $4\pi$ periodic. The evolution of the low-energy spectrum for the quantum wire with increasing spin--orbit energy is shown in Fig.~\ref{fig:STS_TI-limit}. The sinusoidal $2\pi$-periodic dependence of the Majorana states present for $\epsilon_{\rm SO}\simeq B$ becomes a $\pm\cos(\phi/2)$ dispersion with avoided crossings at $\pi$ and $3\pi$ for $\epsilon_{\rm SO}\gg B$. In the limit of large spin--orbit energy, these states reside in the low-momentum subspace and reflect that the topological insulator model displays a topological Josephson effect in this subspace. Similarly, there are also Andreev states in the high-momentum subspace, similar to the ones in $B-\Delta-B$ junctions with the same $\pm E_0\cos(\phi/2)$ dispersion. At large but finite values of the spin--orbit energy, the levels in the low- and high-momentum subspaces mix, resulting in avoided 
crossings at $\phi=\pi,3\pi$ and a $2\pi$-periodic spectrum. The avoided crossings close as the spin--orbit energy diverges, explaining the difference in $\phi$ periodicities of the quantum wire and topological insulator. Finally, the absence of change in the $\theta$ dependence between quantum wire and topological insulator has the same explanation as for $B - \Delta - B$ junctions.

\section{Magnitude of the magneto-Josephson effect}
\label{sec:estimates}

In experiments aimed at detecting the $4\pi$-periodic Josephson effect, a $2\pi$-periodic background current originating from the conventional Josephson effect of the continuum states may mask the signature of the unconventional Josephson current. In the following, we provide quantitative estimates for the $4\pi$- and $2\pi$-periodic contributions to the current and show that the magneto-Josephson effect may be favorable over the conventional Josephson effect with regard to the relative magnitude of $2\pi$- and $4\pi$-periodic currents.

In order to obtain quantitative estimates, we consider a junction with a conventional insulating barrier between two semi-infinite quantum wires in the $B$-dominated phase. In the barrier, we set $B=\Delta=0$ and $\mu=-V_0<0$, so that there are no unconventional $4\pi$-periodic Josephson currents originating from splitting Cooper pairs in the barrier into the two topological superconducting phases on the left and right.\cite{jiang11} Thus, the Josephson currents in this setup are only due to the phase difference $\phi=\phi_l-\phi_r$ or the difference $\theta=\theta_l-\theta_r$ in magnetic-field orientations of the left and right bank. The total energy $E$ (and hence the Josephson current) includes contributions from the above-gap continuum and the Andreev bound states with a $2\pi$-periodic dispersion (jointly refered to below as continuum contribution for brevity) as well as Majorana bound states whose energy is $4\pi$ periodic.

For junctions with a low transmission probability $D\ll 1$, we find the energy of the Majorana states to be
\begin{align}
E_{\rm Majorana}(\phi,\theta)=E_M\cos(\phi/2)\cos[\theta/2+\theta_0(\phi)].
\end{align}
The $\theta$ dependence of the energy involves a phase shift, whereas the $\phi$ dependence is always symmetric with respect to $\phi=0$ (cf.\ the lower panels of Figs.~\ref{fig:TST} and \ref{fig:STS}). The largest energy splitting is given by $E_M\sim \sqrt{D}E_{\rm gap}$, where $E_{\rm gap}$ denotes the magnitude of the gap in the two banks of the junction. The size of the splitting is determined by the single-electron tunneling amplitude $\propto \sqrt{D}$.

\begin{figure}[t]
\centering
\includegraphics[width=.45\textwidth]{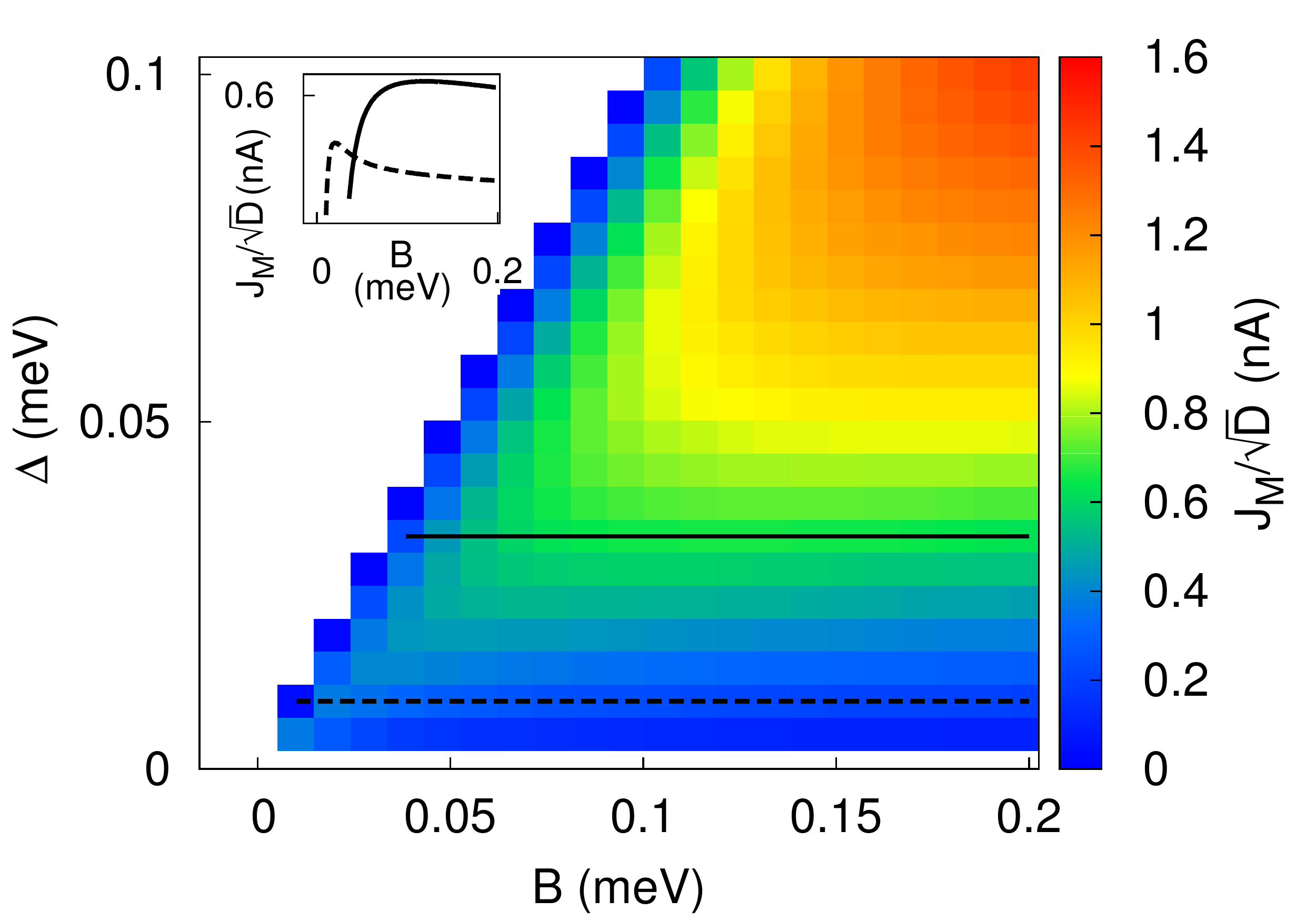}
\caption{Maximum Majorana Josephson current $J_M$ for a $B-I-B$ junction normalized by the normal-state transmission coefficient $\sqrt{D}$ of the junction as a function $B$ and $\Delta$. We set $\mu=0$ in the superconductor, which thus supports a topological phase for $B>\Delta$. Inset: cuts along the dashed and solid lines in the color scale plot. The parameters are $\epsilon_{SO}=0.05$meV, $m=0.015m_e$, $V_0=250$meV, $L_j=3.2$nm.}
\label{fig:num-est_2d}
\end{figure}

The critical current of the junction depends on $\theta$ with a maximum critical current of $J_M=(e/\hbar)E_M$. In Fig.~\ref{fig:num-est_2d}, we show numerical results for $J_M$, normalized by $\sqrt{D}$ to make the results insensitive to detailed properties of the tunnel junction, as a color scale plot. This normalized Majorana current roughly corresponds to $(e/\hbar) E_{\rm gap}$. Thus for a fixed $\Delta$, there is an optimal value of $B$ for which the ratio $J_M/\sqrt{D}$ is maximized.

\begin{figure}[t]
\centering
 \includegraphics[width=.45\textwidth]{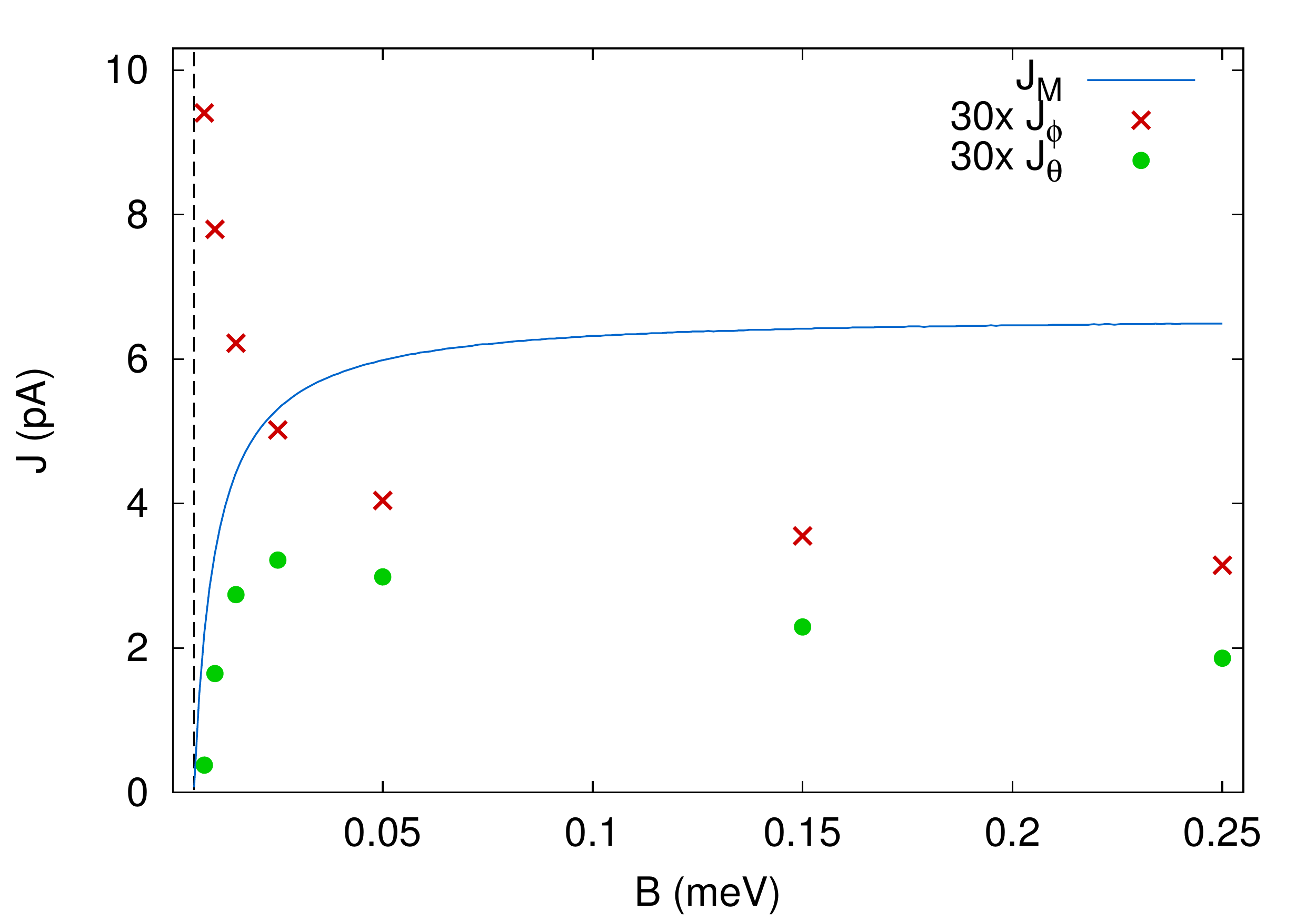} 
\caption{Numerical results for the Majorana ($J_M$) and continuum ($J_\phi$, $J_\theta$) current oscillation amplitudes as a function of B. For presentation the continuum current has been multiplied by 30. The calculations have been performed with the same parameters as in Fig.~\ref{fig:num-est_2d} with $\Delta=5\mu$eV. Transmission probabilities range from $0.1\%$ to $0.4\%$. The vertical dashed line denotes the critical magnetic field.}
\label{fig:num-est_1d}
\end{figure}

In Fig.~\ref{fig:num-est_1d}, we compare the amplitudes of the $4\pi$- and $2\pi$-periodic currents originating from the Majorana and continuum states, respectively. We distinguish between the modulation of the continuum current with $\phi$ and $\theta$. According to our numerical results, the phase dependence of the continuum energy can be well described by
\begin{align}
 E_{\rm cont}(\phi,\theta)=&E_0\cos(\phi)\left\{\alpha +(1-\alpha)\cos\left[\theta+\theta_1(\phi)\right]\right\rbrace\nonumber\\
 &+{\rm const.},
\end{align}
where $\alpha\in[0,1]$ is a parameter-dependent constant. Hence for $\alpha>0$, the amplitude of the oscillations as a function of $\theta$ is smaller than the amplitude of the $\phi$-oscillations. We numerically calculate the largest amplitudes of the Josephson current oscillations as a function of $\phi$ and $\theta$,
\begin{align}
 J_\phi&= eE_0/\hbar\nonumber\\
 J_\theta&=(eE_0/\hbar)(1-\alpha)\label{bulk_current}
\end{align}
and plot them in Fig.~\ref{fig:num-est_1d} as a function of $B$ along with the amplitude of the $4\pi$-periodic Majorana current $J_M$. The latter is much larger than the continuum contribution for a large range of parameters. For a $p$-wave-superconductor junction the $2\pi$-periodic current involves tunneling of Cooper pairs with amplitude $\propto D$,\cite{alicea11} in contrast to single-electron tunneling $\propto \sqrt{D}$ responsible for the $4\pi$-periodic current. Hence in the large $B$ limit, we expect $J_M$ to exceed the $2\pi$-periodic current by a factor of $1/\sqrt{D}$, which is $\sim 20$ for the parameters used in Fig.~\ref{fig:num-est_1d}.

Only very close to the phase transition, when $|B-\Delta|\ll\Delta$, can the continuum current exceed the Majorana contribution. This is consistent with numerical estimates for continuum and Majorana Josephson currents for  the topological insulator edge (\ref{eq:DefH}) in Ref.~\onlinecite{jiang11}, which corresponds to the limit $|B-\Delta|\ll\Delta$ for the quantum wire model (\ref{hamil_QW}).

\begin{figure}[t]
\centering
  \includegraphics[width=.23\textwidth]{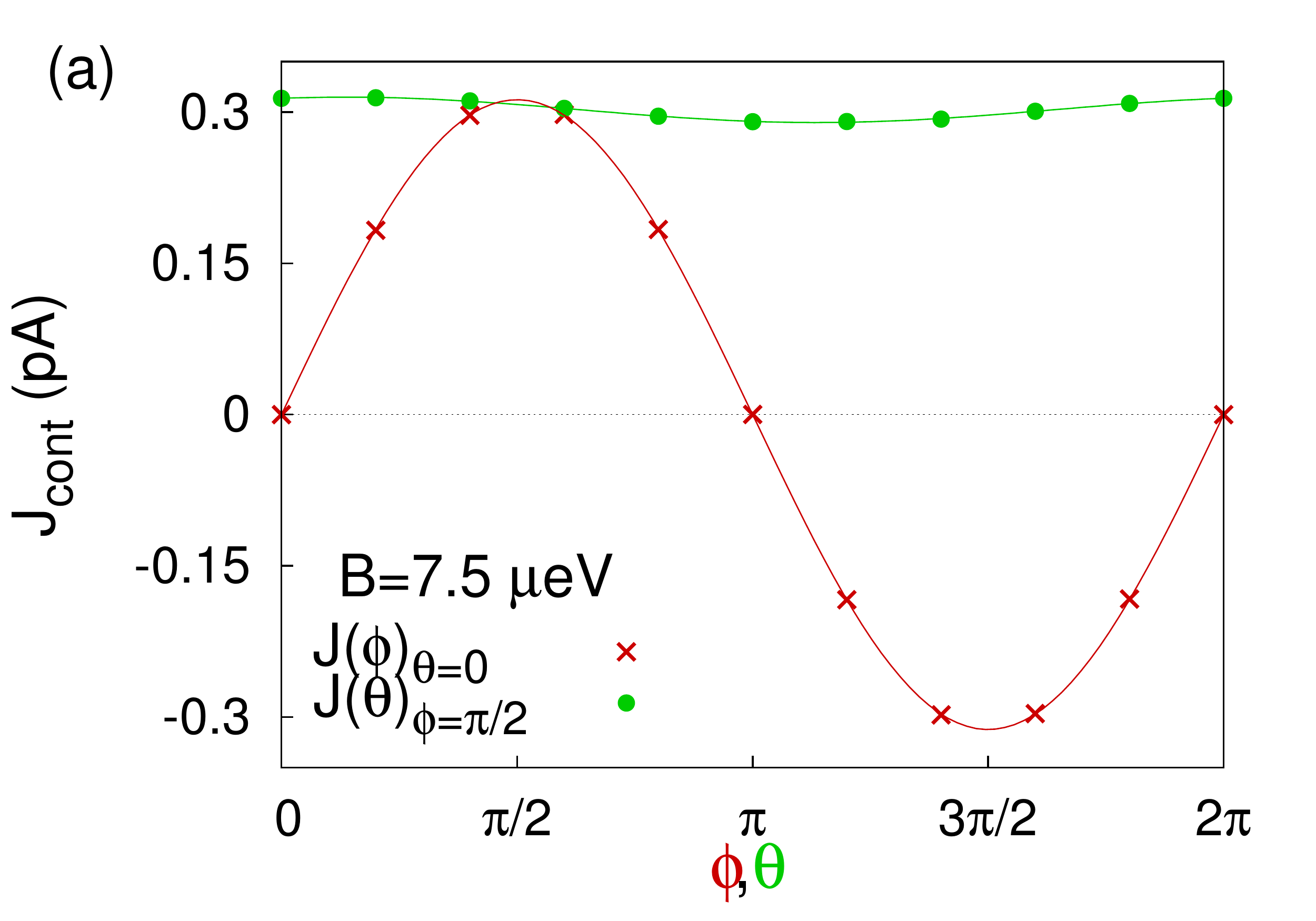}
  \includegraphics[width=.23\textwidth]{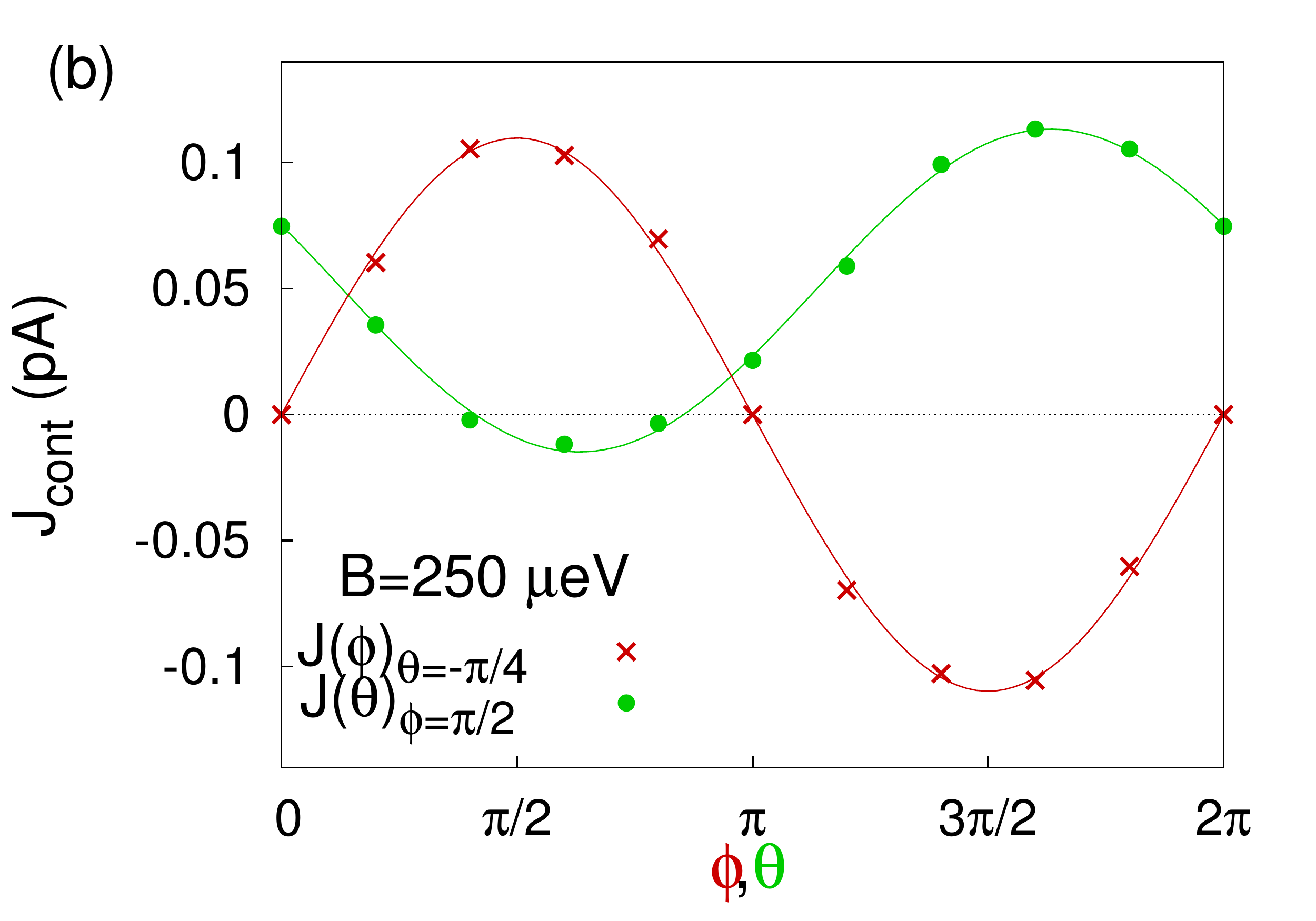}
\caption{Numerical results for the continuum contribution to the Josephson current as a function of $\phi$ (red crosses) and $\theta$ (green dots). The solid lines are fits to the $\phi$-derivative of Eq.~(4), $\partial_\phi E_{\rm cont}(\phi,\theta)$. All parameters are the same as in Fig.~\ref{fig:num-est_1d}. The fixed angle is chosen such that the current is maximized. In this way, the oscillation amplitudes can be used to obtain $J_\phi$ and $J_\theta$ plotted in Fig.~\ref{fig:num-est_1d}. The $\phi$ dependence is simply $J(\phi)=J_\phi \sin(\phi)$. In contrast $J(\theta)$ has a large offset and only a weak $\theta$ dependence for $B\sim\Delta$ [panel (a)]. This corresponds to $\alpha\sim 1$. For $B\gg \Delta$ [panel (b)], the amplitude of $J_\theta$ approaches that of $J_\phi$ and $\alpha$ decreases.}
\label{fig:continuumJ}
\end{figure}

Comparing the Josephson and magneto-Josephson effects, we find in accordance with Eq.~(\ref{bulk_current}) that $J_\phi$ is larger than $J_\theta$, in particular in the regime of small $B$. On the other hand, the amplitude $J_M$ of the Majorana current oscillation is the same for $\phi$ and $\theta$. 
Thus, near the critical magnetic field, the $4\pi$-periodic magneto-Josephson current appears on top of a constant current background with a small $2\pi$-periodic modulation from the continuum states (see Fig.\ \ref{fig:continuumJ}a). This is favorable in experiments to discriminate the $4\pi$-periodic Majorana current from the conventional Josephson current of the continuum, e.g., in the Shapiro-step-like pattern due to the interference of a rotating magnetic field and an $ac$ voltage as described in Ref.~\onlinecite{jiang13}.

\section{Conclusions}
\label{sec:conclusions}

The dedicated experiments\cite{mourik12,das12,churchill13,rokhinson12,lund,harlingen} to detect Majorana bound states in spin--orbit-coupled quantum wires raise the question of how the exotic signatures of Majorana zero modes manifest themselves in such wires. In this work, we explored the magneto-Josephson effect which complements the remarkable Josephson physics that Majorana bound states entail.

Our principal goal was to determine the periodicities of the magneto-Josephson effect for the various domain configuration of the Josephson junction. For junctions made of topological insulator edge states, the $4\pi$-periodic effects emerge in mutually exclusive configurations: the conventional Josephson effect (involving a phase difference of the superconducting order parameter across the junction) appears in the $\Delta-B-\Delta$ domain sequence, while the magneto-Josephson effect requires the complementary structure, $B-\Delta-B$. This indicates that, in a sense, both domain types are
topological for a topological-insulator edge. In contrast, for spin--orbit-coupled wires, we found for both types of Josephson effects that a $4\pi$
periodicity requires the $B-\Delta-B$ configuration.

While we invoked both analytical arguments and numerical analysis to establish this result, it has a natural interpretation. We expect that $4\pi$ periodicity with a parameter which is normally defined between $0$ and $2\pi$ (up to trivial shifts) can only emerge if the parameter pertains to a topological phase. In a spin--orbit-coupled wire, there is no ambiguity as to which phase is topological. A $\Delta$-dominated phase is continuously connected to the vacuum by taking the limit of a large and negative  chemical potential. The $B$-dominated phase, on the other hand, is a topological phase continuously connected to a spinless $p$-wave superconductor. In Sec.\ \ref{analytical}, this argumentation is made explicit using analytical arguments.

Despite these characteristic differences of the magneto-Josephson (as well as Josephson) periodicities between topological insulator edge and
semiconductor quantum wire, both models can be connected explicitly in the limit of large spin--orbit coupling. We exploited this connection in Sec.\
\ref{strongSO} to understand the relation between the Josephson periodicities of the two models.

With a view towards experiments on Majorana Josephson phenomena, we also computed the $4\pi$-periodic magneto-Josephson effect for typical
parameters, and compared its magnitude to that of the more conventional $2\pi$-periodic background. The above-gap continuum of states in the wire
contributes to both the phase-controlled and the magneto-Josephson effect. In the low-transmission regime ($D\ll 1$), we found that both $4\pi$-periodic Josephson effects yield currents of the order of $\frac{e}{\hbar}\sqrt{D} E_{gap}$, with $E_{gap}$ the gap in the two banks of the Josephson junction. In contrast, the conventional effects are suppressed by an additional factor of $\sqrt{D}$.
In order to measure a sizeable $4\pi$-periodic current in experiment, however, it may be necessary to work at large transmission probabilities. In this regime, the exotic and conventional current contributions are of the same magnitude, although the $4\pi$ periodicity is relatively more pronounced for the magneto-Josephson effect.

Josephson-related phenomena in spin--orbit-coupled wires can become more complex when considering, e.g., $ac$ modulations and Shapiro steps. These may require one to take accurate account of the complicated spectrum of the quantum wires. Moreover, when the energy gap in the middle domain is not too large, additional Andreev bound states could be present, which contribute to the Josephson effect. In this work we refrain from discussing these topics as well as more complicated setups such as three-leg Josephson effects to keep the presentation concise. Nonetheless, these aspects may prove important (and maybe even beneficial) in experiments, and present interesting avenues for future research. 

\acknowledgments

We thank Arbel Haim for discussions and are grateful for support from the Helmholtz Virtual Institute ``New states of matter and their excitations," SPP1285 (DFG), NSF grant DMR-1055522, ISF, BSF, a TAMU-WIS grant, NBRPC (973 program) grant 2011CBA00300 (2011CBA00301), the Alfred P. Sloan Foundation, the Packard Foundation, the Humboldt Foundation, the Minerva Foundation, the Sherman Fairchild Foundation, the Lee A. DuBridge Foundation, the Moore Foundation funded CEQS, the Institute for Quantum Information and Matter (IQIM), NSF Physics Frontiers Center with support of the Gordon and Betty Moore Foundation, and the Studienstiftung des dt.\ Volkes.\\

\appendix

\section{Quantum wire in the limit of large spin--orbit coupling}
\label{appendix}

In Sec.\ \ref{strongSO}, we relied strongly on the statement that in the limit of strong spin--orbit coupling, the quantum-wire model (\ref{hamil_QW}) reduces at low energies to a combination of the topological-insulator-like low-momentum subspace and a spinless-$p$-wave-superconductor-like high-momentum subspace. In this appendix, we provide an explicit justification for this statement. 

The statement is evident for the low-momentum subspace, so we will not consider it further. For $\mu=0$, the Fermi points are located at $p=0$ (low momentum) as well as $p=\pm p_F$ with $p_F = 2mu$ (high momentum). As the spin--orbit coupling (or equivalently, $m$) increases, $p_F$ becomes large and so does the effective spin--orbit field in the high-momentum subspace. Thus, in this limit, we can first diagonalize the Hamiltonian in the absence of the induced pairing $\Delta$ and then treat the latter perturbatively. To do so, we perform the unitary transformation 
\begin{equation}
  {\cal U} = \exp\{ i\alpha \sigma_y \tau_z /2  \} \exp\{ i\theta \sigma_z /2  \}
\end{equation}
on the wire Hamiltonian (\ref{hamil_QW}) with $\Delta = 0$. If we choose $\alpha$ such that $\tan\alpha= B/up$, the rotated Hamiltonian takes the form
\begin{equation}
   {\mathcal H}_0 = \left(\frac{{\hat p}^2}{2m} + \sqrt{(up)^2 + B^2} \sigma_z\right)\tau_z.
\end{equation}
The low-energy subspace at $p=\pm p_F$ is formed by the bands for which $\sigma_z$ takes the value $-1$. We now reintroduce the pairing term $\Delta\tau_x$ and apply the transformation $\mathcal{U}$ to it.
The projection of ${\mathcal H}_0$ onto the lower bands yields
\begin{align}
 {\mathcal H}_{\rm eff}=\left(\frac{{\hat p}^2}{2m} - \sqrt{(up)^2 + B^2} \right)\tau_z+ \frac{up\Delta}{\sqrt{(up)^2 + B^2}}\tau_x.
\end{align}
The condition $\epsilon_{SO}\gg\Delta$ guarantees that we can neglect the coupling to high-energy degrees of freedom near $\pm p_F$. Linearizing around the Fermi momenta and using $|up|\sim\epsilon_{SO}\gg B$ the effective Hamiltonian takes the form
\begin{align}
 {\mathcal H}_{\rm eff}=u\left(|p|-p_F \right)\tau_z+ {\rm sign}(p)\Delta\tau_x.
\end{align}
This describes a spinless $p$-wave superconductor.

\end{document}